\documentclass[twocolumn,preprintnumbers,amsmath,amssymb,aps,prl,reprint]{revtex4}
\usepackage{graphicx}
\begin{document}

\title{
Thermal Creep and the Skyrmion Hall Angle in Driven Skyrmion Crystals     
} 
\author{
C. Reichhardt and C. J. O. Reichhardt 
} 
\affiliation{
Theoretical Division and Center for Nonlinear Studies,
Los Alamos National Laboratory, Los Alamos, New Mexico 87545, USA\\ 
} 

\date{\today}
\begin{abstract}
We numerically examine thermal effects on the skyrmion Hall angle 
for driven skyrmions interacting with quenched disorder. 
We identify a creep regime in which  
motion occurs via intermittent jumps between pinned and flowing states.
Here the skyrmion Hall angle is zero since the
skyrmions have time to relax into equilibrium positions in the pinning sites,
eliminating the side-jump motion induced by the Magnus force.
At higher drives we find a crossover to a viscous flow regime 
where the skyrmion Hall angle is finite and increases with increasing drive or
temperature.
Our results are in agreement 
with recent experiments which also
show a regime of
finite skyrmion velocity with zero skyrmion Hall angle
crossing over to a viscous flow regime with a skyrmion Hall angle that increases
with drive.
\end{abstract}
\maketitle

A wide range of systems with quenched disorder 
exhibit depinning phenomena under an external drive  
\cite{1,2}, including sliding charge density waves \cite{3}, 
vortices in type-II superconductors \cite{4},
colloids \cite{5,6} and magnetic domain walls \cite{7}.
At zero temperature, these systems have a well defined depinning 
threshold $F_{c}$ below which motion does not occur,
while at finite temperature,
a finite velocity arises
at much lower drives due to thermal creep effects  \cite{1}.
Another system that exhibits 
depinning phenomena is skyrmions in chiral magnets \cite{8,9,10}. 
Skyrmions are topological particle-like magnetic objects that can  
form lattices and be driven by
an applied current
or by other methods 
\cite{8,11,12}.  
Skyrmion velocity-force relations can be obtained  
by transport measures \cite{9,13} and direct imaging \cite{14,15,16,17}.
The topology of the skyrmions makes them unique among systems that
exhibit depinning since the skyrmion dynamics
is strongly influenced by 
the Magnus force, which generates a velocity
component perpendicular to the forces experienced by the skyrmion 
\cite{8}.

One 
 consequence of the Magnus force is that skyrmions move at an angle,
 called the skyrmion Hall angle $\theta_{sk}$, with respect 
 to an external driving force
 \cite{8,16,17,18}.
 The intrinsic value  of $\theta_{sk}$ is determined by the ratio of the
 Magnus term to the damping
 term, and in the absence of quenched disorder,
 $\theta_{sk}$ is a constant. 
Particle-based
simulations of moving skyrmions interacting with quenched disorder have 
shown that 
$\theta_{sk}$ can vary,
starting at $\theta_{sk}=0$ at depinning, increasing with increasing velocity,
and
saturating to the defect-free intrinsic value 
at higher drives \cite{19,20,21,22}.
This effect arises due to the
Magnus-induced side jump experienced by the skyrmions
when they move through a pinning site
\cite{20,23}. 
Under rapid skyrmion motion,
the magnitude of the side jumps is reduced
since the skyrmion does 
not have time to respond fully to the pinning potential
before exiting the pinning site.
Continuum-based simulations
reveal
a similar drive dependence of $\theta_{sk}$
when disorder is present,
with $\theta_{sk}$ remaining constant in the absence of disorder
\cite{24,25}.
In imaging experiments  
of skyrmions under an external drive,
$\theta_{sk}=0$ at low drives just above depinning,
but as the drive increases, $\theta_{sk}$ increases linearly
until it reaches a saturation at 
high drives close to the predicted intrinsic value \cite{16}. 
In other experiments which
show a similar drive dependence of $\theta_{sk}$,
it was argued that in addition to the pinning interactions,
excitation of internal modes of the skyrmions or
a change in skyrmion size with driving
can modify the skyrmion Hall angle
\cite{17}.
In more recent work
it was claimed that both of
these effects
contribute to changes in $\theta_{sk}$ \cite{26}.
Far less is known about how thermal creep affects
skyrmion motion or the value of $\theta_{sk}$. 
Thermal effects 
are
relevant since many materials support skyrmions at 
room temperature \cite{14,15,16,18,27,28}.
Additionally, in recent experiments,
thermally induced Brownian motion of skyrmions has been
observed directly
in the 
absence of an external drive \cite{29}. 

In this work we examine the thermal creep of skyrmions and
its effect on
$\theta_{sk}$ 
using a particle-based skyrmion model. 
Thermal motion is most significant
near the depinning threshold where the drive is small, 
in a regime where the effect of the skyrmion shape
on $\theta_{sk}$,
which has been proposed to be important at high drives \cite{17}, is minimal. 
We find that
the depinning threshold decreases with increasing temperature
and that a creep regime appears which is
characterized by 
skyrmions jumping or even avalanching
between pinned and moving states.
Within the creep regime,
$\theta_{sk} = 0$,
while at higher drives there is a crossover
to viscous flow where
the skyrmions are always moving
and the value of $\theta_{sk}$ becomes finite and
increases with drive.
We note that
recent
particle-based simulations
have examined thermal effects in skyrmions, 
but these studies focused on the aging
dynamics in the absence of a drive \cite{30}.  

{\it Simulation-- } 
We consider  
a two-dimensional system of rigid skyrmions
with periodic boundary conditions in the $x$- and $y$-directions.
The skyrmion dynamics are modeled using a modified version 
of the Thiele equation \cite{19,20,21,31},
and we include Langevin dynamics 
that are similar to those employed previously 
to model thermal relaxation effects in skyrmions \cite{30}. 
The equation of motion of skyrmion $i$ is 
$\alpha_d {\bf v}_{i} + \alpha_m {\hat z} \times {\bf v}_{i} =
{\bf F}^{ss}_{i} + {\bf F}^{p}_{i} + {\bf F}^{D} + {\bf F}^{T}_{i},$ 
where
${\bf v}_{i} = {d {\bf r}_{i}}/{dt}$
is the skyrmion velocity 
and $\alpha_d$  
and $\alpha_m$ are the damping and Magnus terms,
respectively. The 
intrinsic skyrmion Hall 
angle is given by $\theta^{in}_{sk} = \tan^{-1}(\alpha_{m}/\alpha_{d})$.
We fix $\alpha_m/\alpha_d=1.0$ so that $\theta^{in}_{sk}=45^\circ$,
which is close to the value in recent experiments \cite{16,17}.
The skyrmion-skyrmion interaction force
${\bf F}_{i}^{ss} = \sum^{N}_{j \neq i}K_{1}(r_{ij}){\hat {\bf r}_{ij}}$, 
where $K_{1}$ is the modified Bessel function, $r_{ij} = |{\bf r}_{i} - {\bf r}_{j}|$ 
is the distance between skyrmions $i$ and $j$,
${\hat {\bf r}}_{ij}=({\bf r}_{i}-{\bf r}_j)/r_{ij}$,
and $N$ is the number of skyrmions in the sample \cite{19,31,32}.   
The pinning force ${\bf F}^{sp}_i = \sum_{j=1}^{N_p}(F_{p}/r_{p})({\bf r}_{i} -{\bf r}_{p}^j)\Theta(r_{p} -|{\bf r}_{i} - {\bf r}_{p}^j|)$ arises from randomly placed non-overlapping parabolic
traps
of radius $r_{p} = 0.15$ at locations ${\bf r}_p^j$
that can exert a maximum pinning force of $F_{p}=0.03$
on a skyrmion.  Here $\Theta$ is the Heaviside step function.
The driving force from an applied current is ${\bf F}^{D} = F_{D}{\hat {\bf x}}$,
and  we measure the average velocity
per skyrmion
both parallel, $\langle V_{||}\rangle = N^{-1}\sum^{N}_{i=1} {\bf v}_{i} \cdot {\bf \hat{x}}$,
and perpendicular,
$\langle V_{\perp}\rangle = N^{-1}\sum^{N}_{i=1} {\bf v}_{i} \cdot {\bf \hat{y}}$,
to the drive.
The skyrmion Hall angle is given by
$\theta_{sk}=\tan^{-1}(\langle V_{\perp}\rangle/\langle V_{||}\rangle)$.
The stochastic thermal force ${\bf F}^{T}_{i}$ has the properties
$\langle {\bf F}^T_{i}\rangle = 0$
and
$\langle {\bf F}^T_{i}(t){\bf F}^{T}_{j}(t^\prime)\rangle =  2\eta k_{B}T\delta_{ij}\delta(t -t^\prime)$.
Here we set $\eta = 1.0$ and $k_{B}= 1.0$.
We use a skyrmion density of $n_s=N/L^2=0.16$, where the sample is of size
$L \times L$ with $L=36$.  We fix the pinning density at $n_p=N_p/L^2=0.2$.

\begin{figure}
\includegraphics[width=\columnwidth]{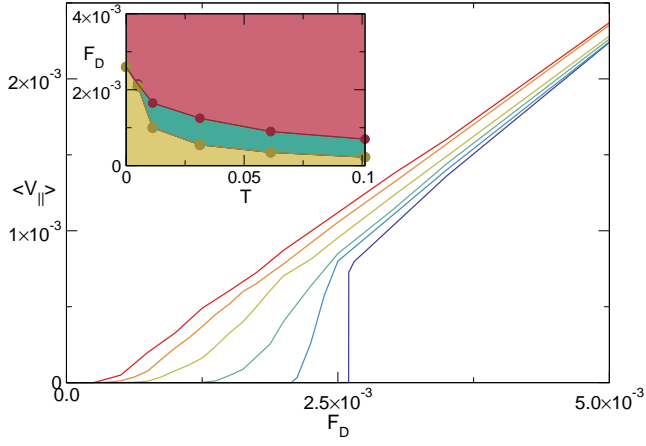}
\caption{The time average skyrmion velocity in the direction parallel to the drive,
  $\langle V_{||}\rangle$, vs $F_{D}$
  at  temperatures
  $T = 0.0$ (dark blue),
  $0.005$ (light blue),
  $0.01125$ (dark green),
  $0.03125$ (light green),
  $0.06125$ (orange),
  and $0.1025$ (red).
  As $T$ increases,
  there is a decrease in the depinning threshold $F_{c}$ and the behavior
  of $\langle V_{||}\rangle$
  becomes increasingly nonlinear.
  Inset:  Dynamic phase diagram as a function of $F_{D}$ vs $T$
  highlighting the pinned (yellow),
  creep (green), and viscous flow (pink) regimes.
}
\label{fig:1}
\end{figure}

{\it Results--}
In Fig.~\ref{fig:1} we plot $\langle V_{||}\rangle$ versus $F_{D}$
for temperatures ranging
from $T=0$ to $T=0.1025$.
At $T = 0$, there is a 
sharp depinning threshold
at a critical depinning force of
$F_{c} = 2.6 \times 10^{-3}$.
As $T$ increases, $F_{c}$ decreases
and the regime of nonlinear behavior of $\langle V_{||}\rangle$ grows in extent.
We measure
the evolution of $\theta_{sk}$
at each temperature as a function of drive.
Previous work at $T = 0$
showed that $\theta_{sk}$ is zero only in the
pinned phase, and that $\theta_{sk}$ increases
with increasing $F_D$ for $F_{D}/F_{c} > 1.0$ \cite{20,21,22}.
At finite temperature
we find 
an extended creep regime
in which
$\langle V_{||}\rangle > 0$
while $\langle V_{\perp}\rangle = 0$, giving $\theta_{sk} = 0$
even though the skyrmion velocity is finite.
Within the creep regime,
the skyrmion motion is intermittent and consists of jumps
between moving and pinned states.
At higher drives, the creep regime 
transitions into a  viscous flow phase in which
the skyrmions are always in motion and
$\theta_{sk}$
increases with $F_D$.
Based on
measurements of $\langle V_{||}\rangle$, $\langle V_{\perp}\rangle$,
$\theta_{sk}$, and 
histograms of the skyrmion velocities,
we construct a dynamic phase diagram as a function of
$F_D$ versus $T$ highlighting 
the pinned phase with $\langle V_{||}\rangle = \langle V_{\perp}\rangle = 0$,
the creep phase with $\langle V_{||}\rangle > 0$  and
$\langle V_{\perp}\rangle = 0$, and the viscous flow 
regime
with $\langle V_{||}\rangle > 0$ and $\langle V_{\perp}\rangle > 0$,
as shown in the inset of Fig.~\ref{fig:1}.

\begin{figure}
\includegraphics[width=\columnwidth]{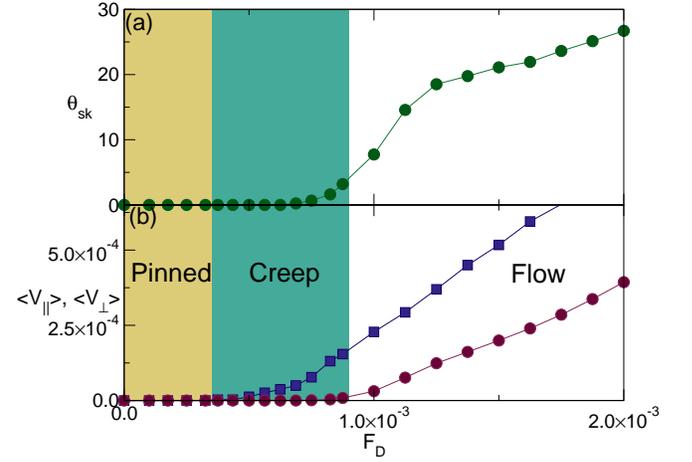}
\caption{(a) $\theta_{sk}$ vs $F_{D}$
  at $T = 0.06125$.
  (b) $\langle V_{||}\rangle$ (blue squares) and $\langle V_{\perp}\rangle$ (red circles)
  vs $F_{D}$ in the same system.
  Here there is a pinned phase, a creep phase in which $\theta_{sk}$ is close to zero, and 
a flowing phase.   
}
\label{fig:2}
\end{figure}

\begin{figure}
\includegraphics[width=\columnwidth]{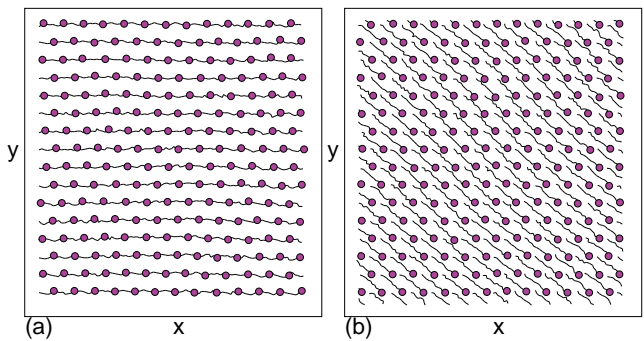}
\caption{ 
  The skyrmion positions (dots) and trajectories (lines)
  for the system in Fig.~\ref{fig:2} at $T = 0.06125$.
  (a) The creep phase at
  $F_{D} = 7.5\times 10^{-4}$,
  where $\theta_{sk} = 0$ and the skyrmions move in the direction of drive. 
  (b) The flow phase at
  $F_{D} = 3.5 \times 10^{-3}$,
  where
  $\theta_{sk} = 35^\circ$.  
}
\label{fig:3}
\end{figure}

In Fig.~\ref{fig:2}(a) we plot
$\theta_{sk}$
versus $F_{D}$
at $T=0.06125$, 
and in Fig.~\ref{fig:2}(b) we show
the corresponding $\langle V_{||}\rangle$ and $\langle V_{\perp}\rangle$ versus
$F_D$ curves.
In the pinned phase,
both $\langle V_{||}\rangle$ and $\langle V_{\perp}\rangle$ are zero,
while in the creep regime, 
$\langle V_{||}\rangle$ is finite
but $\langle V_{\perp}\rangle$ is zero or nearly zero,
so that
$\theta_{sk}$ is zero or close to zero.
In the flow regime, both velocity components increase and $\theta_{sk}$
rises from zero.
When the system first enters the flow regime,
the increase in $\theta_{sk}$ with $F_D$ is nonlinear,
but there is a crossover to a linear increase in $\theta_{sk}$ at higher
drives.
Above the range of drives illustrated in Fig.~\ref{fig:2},
$\theta_{sk}$ saturates to the intrinsic value of $\theta_{sk}^{in}=45^\circ$.
The pinned, creep, and flow behavior
of $\langle V_{||}\rangle$, $\langle V_{\perp}\rangle$,
and $\theta_{sk}$
is very similar to
the observations in
imaging experiments of the room temperature motion of skyrmions \cite{16}.  
In Fig.~\ref{fig:3}(a)
we show the skyrmion positions and trajectories for the system in Fig.~\ref{fig:2}
in the creep regime at $T=0.06125$ and
$F_{D} = 7.5\times 10^{-4}$,
where the skyrmions move in the direction of drive.
At the same temperature but at a higher drive of
$F_D=3.5\times 10^{-3}$
in Fig.~\ref{fig:3}(b),
the skyrmions are in the flow regime and move at an angle of
$\theta_{sk}=35^\circ$ with respect to the drive.  
 
\begin{figure}
\includegraphics[width=\columnwidth]{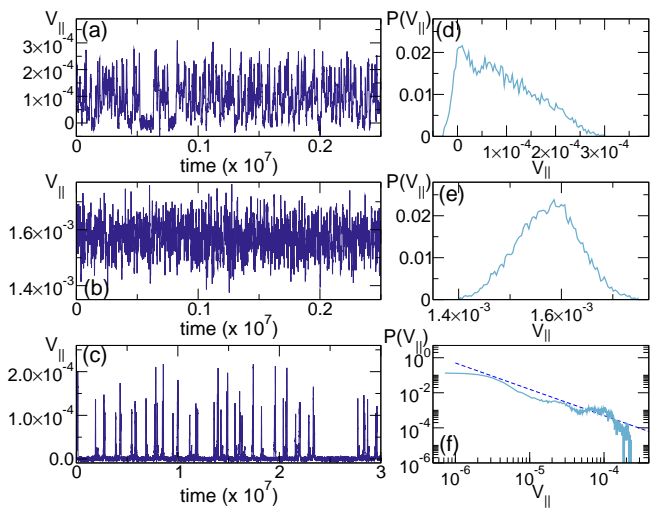}
\caption{(a) Time series of $V_{||}(t)$ in units of $10^7$ simulation time steps
  in the creep regime at
  $F_{D} = 7.5\times 10^{-4}$
  and $T = 0.06125$.
  (b)
  The corresponding velocity histogram
  $P(V_{||})$.
  (c) $V_{||}(t)$ and (d) $P(V_{||})$
  in the flow regime at
  $F_{D} = 3.5\times 10^{-3}$ and $T=0.06125$,
  where $\langle V_{||}\rangle > 0$ and $\langle V_{\perp}\rangle > 0$.
  (e) $V_{||}(t)$
  in the creep regime at
  $F_{D} = 1.5\times 10^{-3}$
  and $T =  0.01125$, where
  motion occurs in the form of avalanches.
  (f) The corresponding histogram $P(V_{||})$,
  where the dashed line is a power law fit
  to $P(V_{||}) \propto V_{||}^{-\alpha}$ with $\alpha = 1.5$.
}
\label{fig:4}
\end{figure}

The skyrmion motion within the creep phase
has a stop-start character, where a skyrmion that is trapped by a
pin escapes from the pin and undergoes a brief interval of motion before
becoming trapped by an adjacent pin.
While trapped, the skyrmion spirals toward the equilibrium
location in the pinning site at which the pinning and driving forces are
exactly balanced.
In Fig.~\ref{fig:4}(a,b) we plot a velocity time series $V_{||}(t)$ and corresponding
velocity histogram $P(V_{||})$
in the creep regime at
$F_{D} = 7.5\times 10^{-4}$
and $T = 0.06125$.
There is a peak in $P(V_{||})$ at $V_{||}=0$, indicating that the skyrmions
spend the largest portion of their time trapped in pinning sites.
In the flow regime,
shown in Fig.~\ref{fig:4}(c,d)
at
$F_{D} = 3.5\times 10^{-3}$
and $T = 0.06125$,
the skyrmions are constantly
flowing and $V_{||} > 0$ at all times.

In the creep regime,
$\theta_{sk} = 0^\circ$
because the Magnus force is a dynamical quantity that
can act only when the skyrmions are driven out of equilibrium.
If there is sufficient time between jumps of the skyrmion from one
pinning site to another,
the skyrmion
gradually spirals to a
position in the pinning 
site where the pinning and drive forces are balanced.
The location of
this point is independent of the Magnus term, so
over long times, a skyrmion in the strictly overdamped limit of
$\alpha_{m} = 0$
settles into
the same position as  a skyrmion with $\alpha_{m} > 0$.
On the other hand, if the skyrmions are continuously moving over the pinning sites,
they experience a side jump motion generated by the Magnus force that
pushes them away from the equilibrium force balance point of the pin,
producing a finite value of $\theta_{sk}$ with a magnitude that
depends on the strength of the Magnus force.
As $F_D$ increases and the skyrmions move faster, the perturbation of the
Magnus force by the pinning is reduced and
$\theta_{sk}$ increases until it reaches the intrinsic value $\theta_{sk}=\theta_{sk}^{in}$.
For increasing temperature $T$ at fixed
$F_{D}$,
thermal fluctuations smear out the effectiveness of the
pinning,
permitting the skyrmions
to move continuously and
causing $\theta_{sk}$
to increase.

If the temperature is lowered, 
within the creep regime the
skyrmions spend an even larger fraction of their time trapped in pinning sites,
the motion becomes increasingly intermittent,
and avalanches dominate the behavior,
as illustrated in the plot of $V_{||}(t)$ in Fig.~\ref{fig:4}(e)
at $F_{D} = 1.5\times 10^{-3}$
and $T = 0.01125$.
In Fig.~\ref{fig:4}(f) we show $P(V_{||})$ in the avalanche regime on a log-log
scale,
where the dashed line is a power law fit to
$P(V_{||}) \propto V_{||}^{-\alpha}$ with $\alpha = 1.5$.
In previous work, we
examined
the avalanche behavior
just at the depinning threshold for a $T=0$ system, 
and found that various quantities such
as the avalanche size and duration are power law distributed
with exponents similar to that in Fig.~\ref{fig:4}(f) \cite{32}. 
Here we find that under finite temperature
in the creep regime, the thermal fluctuations lower $F_{c}$ 
but the system can still
exhibit critical behavior near the depinning threshold.
These results indicate that in certain cases, thermally induced skyrmion
avalanches can be realized with a fixed current.

\begin{figure}
\includegraphics[width=\columnwidth]{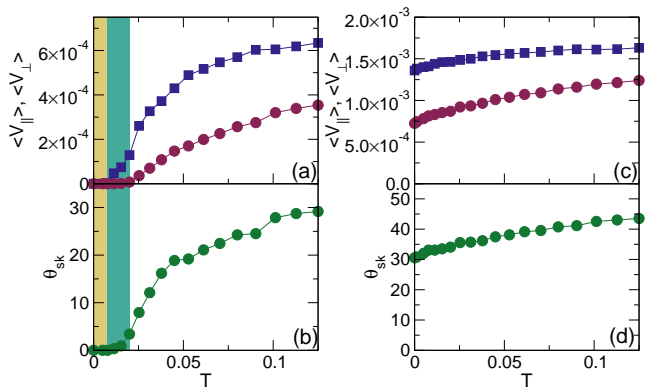}
\caption{(a) $\langle V_{||}\rangle$ (blue squares) and $\langle V_{\perp}\rangle$
  (red circles) vs $T$
  at
  $F_{D} = 1.5 \times 10^{-3}$. 
  (b) The corresponding
  $\theta_{sk}$ vs $T$.
  Yellow shading indicates the pinned regime, green shading is the creep state,
  and the unshaded area is in the flowing state.
  (c) $\langle V_{||}\rangle$ (blue squares) and $\langle V_{\perp}\rangle$
  (red circles) vs $T$ for the same system in the
  flowing state at $F_{D} = 3.5 \times 10^{-3}$.
  (d) The corresponding
  $\theta_{sk}$ vs $T$ curve shows a linear increase in $\theta_{sk}$ with increasing
  $T$ up to the intrinsic value of $\theta_{sk}^{in}=45^\circ$.
}
\label{fig:5}
\end{figure}

In Fig.~\ref{fig:5}
we illustrate the temperature dependence of $\theta_{sk}$ at a fixed $F_{D}$ in two
regimes: 
$F_{D} > F_{c}$ where the skyrmions are always moving,
and $F_{D} < F_{c}$ where the skyrmions spend most of their time pinned. 
In Fig.~\ref{fig:5}(a) we plot $\langle V_{||}\rangle$ and
$\langle V_{\perp}\rangle$
versus $T$
at
$F_{D} = 1.5 \times 10^{-3}$, a drive that is lower than the
$T = 0$ depinning threshold of $F_{c}=2.6 \times 10^{-3}$,
while in Fig.~\ref{fig:5}(b) we plot the corresponding
$\theta_{sk}$ versus $T$.
We find a pinned state
for $T > 0.00725$
and a creep state
for $0.00725 \leq T < 0.02$.
For $T \geq 0.02$, the system is in a flowing phase and
$\theta_{sk}$ increases with increasing $T$.
In Fig.~\ref{fig:5}(c) we show $\langle V_{||}\rangle$ and $\langle V_{\perp}\rangle$
versus $T$ at
$F_{D} = 3.5 \times 10^{-3}$,
a drive that is higher than the $T=0$ value of $F_c$, and in Fig.~\ref{fig:5}(d) we plot
the corresponding $\theta_{sk}$ versus $T$.
At this drive, the skyrmions are in the flowing state for all values of $T$.
The skyrmion Hall angle $\theta_{sk}=31^{\circ}$ at $T=0$ and
increases roughly linearly with increasing temperature before saturating
at the intrinsic value of $\theta_{sk}^{in}=45^\circ$ at high temperatures.
This indicates that a temperature dependence of $\theta_{sk}$ persists
even in
the viscous flow phase. 
We note that in the absence of pinning, $\theta_{sk}$ is independent of both
$F_{D}$ and $T$ within the particle model.
In continuum simulations performed without pinning,
$\theta_{sk}$ was also shown to be
independent of the drive amplitude \cite{24}.

{\it Summary---} 
We have examined the effect of temperature on skyrmion creep and motion
using a particle based model.
As temperature increases, we find that the depinning threshold
drops and a nonlinear creep regime appears at low drives.
Within the creep 
regime, the skyrmion Hall angle is zero and the skyrmions spend most
of their time trapped in pinning sites,
making occasional hops from one pinning site to another.
While trapped, the skyrmions have enough time to spiral to the equilibrium
position in the pin at which the drive and pinning forces are balanced,
eliminating the Magnus-induced side-jump motion that would give a finite value
for the skyrmion Hall angle.
In the flowing regime, the skyrmions are always moving  
and the skyrmion Hall angle increases linearly with increasing drive.
At low temperatures in 
the creep regime,
we find that
skyrmion motion occurs in the form of avalanches,
and that the skyrmion velocities are
power law distributed.
The effect of temperature on the
skyrmion Hall angle is most pronounced at small drives
where thermally-induced depinning occurs;
however, even within the flowing phase,
the skyrmion Hall angle increases with increasing temperature.   

\begin{acknowledgments}
We gratefully acknowledge the support of the U.S. Department of
Energy through the LANL/LDRD program for this work.
This work was carried out under the auspices of the 
NNSA of the 
U.S. DoE
at 
LANL
under Contract No.
DE-AC52-06NA25396.
\end{acknowledgments}

\end{document}